\documentclass[showpacs,preprintnumbers,amsmath,amssymb]{revtex4}
\usepackage{amsmath,amssymb,graphics,epsfig,subfigure}
\usepackage{color}

\begin{document}
\renewcommand{\baselinestretch}{1.3}

\title{Probing the relationship between the null geodesics and thermodynamic phase transition for rotating Kerr-AdS black holes}

\author{Shao-Wen Wei \footnote{weishw@lzu.edu.cn},
        Yu-Xiao Liu \footnote{liuyx@lzu.edu.cn},
        Yong-Qiang Wang \footnote{yqwang@lzu.edu.cn}}

\affiliation{Institute of Theoretical Physics $\&$ Research Center of Gravitation, Lanzhou University, Lanzhou 730000, People's Republic of China}

\begin{abstract}
In this paper, we aim to examine the relationship between the unstable circular photon orbit and the thermodynamic phase transition for a rotating Kerr-AdS black hole. On one side, we give a brief review of the phase transition for the Kerr-AdS black hole. The coexistence curve and the metastable curve corresponding to the phase transition are clearly shown. On the other side, we calculate the radius and the angular momentum of the unstable circular orbits for a photon by analyzing the effective potential. Then combining these two sides, we find the following results. i) The radius and the angular momentum of the unstable circular photon orbits demonstrate the non-monotonic behaviors when the thermodynamic phase transition takes place. So from the behavior of the circular orbit, one can determine whether there exists a thermodynamic phase transition. ii) The difference of the radius or the angular momentum for the coexistence small and large black holes can be treated as an order parameter to describe the phase transition. And near the critical point, it has a critical exponent of $\frac{1}{2}$. iii) The temperature and pressure corresponding to the extremal points of the radius or the angular momentum of the unstable circular photon orbit completely agree with that of the metastable curves from the thermodynamic side. Thus, these results confirm the relationship between the geodesics and thermodynamic phase transition for the Kerr-AdS black hole. Therefore, on one hand, we are allowed to probe the thermodynamic phase transition from the gravity side. On the other hand, the signature of the strong gravitational effect can also be revealed from the black hole thermodynamics.
\end{abstract}

\keywords{Black holes, critical phenomena, phase structure, circular orbit}

\pacs{04.70.Dy, 04.50.Gh, 04.25.-g}

\maketitle

\section{Introduction}

The study of the black hole thermodynamics continues to be an intriguing and important topic in general relativity. It was found that, in asymptotically flat space, the Schwarzschild black hole always has negative heat capacity, and thus, it is thermodynamically unstable. While in anti-de Sitter (AdS) space, there exists a thermodynamically stable large black hole phase. In particular, the Hawking-Page phase transition between the stable large black hole phase and the thermal gas phase was found \cite{Hawking}, and it was explained as the confinement/deconfinement phase transition of a gauge field \cite{Witten2} in the view of the AdS/CFT correspondence \cite{Maldacena,Gubser,Witten}. Motivated by this correspondence, black hole thermodynamics and phase transition have gained a lot of attention.

After fixing the cosmological constant, it was also found that there is small-large black hole phase transition in the charged or rotating AdS black hole backgrounds~\cite{Chamblin,Chamblin2,Caldarelli,Roychowdhury,Banerjee}. Recently, treating the cosmological constant as a new thermodynamic variable \cite{Kastor,Dolan00,Cvetic}, the small-large black hole phase transition was examined in Ref. \cite{Kubiznak}. The precise analogy between the small-large black hole phase transition and the liquid-gas phase transition was established. Subsequently, this van der Waals (VdW)-like phase transition was found in other black hole backgrounds. Moreover, some new phase structures were observed, such as the reentrant phase transition, isolated critical point, triple point, and superfluid black hole phase  \cite{Gunasekaran,Altamirano,Mann,Frassino,Wei0,Kostouki,Wei1,Hennigar,ZouYue,
Hendi,Hendi2,Hendi3,Hendi4,Momeni,Chakraborty,Weisw} (for a recent review see \cite{Teo} and references therein).

For a long time, it was expected that the thermodynamic phase transition of the black hole has some observational signatures, such as the quasinormal modes (QNMs), which can be detected from astronomical observations. These dynamical perturbations provide a possible way to uncover the phase transition. In Ref. \cite{Pan}, the authors proposed that the QNM frequencies of the charged Reissner-Nordstr\"{o}m black holes start to get a spiral-like shape in the complex $\omega$ plane at the Davie's phase transition point, where the heat capacity of the black hole is singular. However, this relation may not be so accurate to describe the phase transition \cite{BertiBerti}. Nevertheless, such a non-trivial relation between the dynamics and  thermodynamics of the black hole was confirmed in Ref. \cite{He}. Other studies on the Davie's phase transition and QNMs of the perturbations can be found in Refs. \cite{Koutsoumbas,Shen,Papantonopoulos,Rao,Myung}. Very recently, it was found in Ref. \cite{Liu} that there exists a dramatic change in the slope of the QNMs among the first-order VdW-like black hole phase transition observed in Ref. \cite{Kubiznak}. This new test was extended to other AdS black hole backgrounds \cite{Mahapatra,Chabab,Zou2,Prasia,www,ZengZeng}. These results imply that the phase transition information can be reflected by the QNMs of the dynamical perturbations.

Considering the above results, we conjecture that there should exist a relationship between the phase transition and gravitational physics for the black hole system, and such a relationship is worth to be tested. The research results in Refs. \cite{Cvetic2,Cvetic3,Tang} imply that the presence of photon orbits signals a possible York-Hawking-Page type phase transition or instability of spacetimes.

However, a clear relationship between the phase transition and gravity is still needed to be constructed. As we know, the unstable circular photon orbit plays an important role in the strong gravitational effects \cite{Cardoso,Stefanov,WeiLiuGuo,HodHod,WeiLiu3,Raffaelli,Franzin,Stuchlik}, so the first step toward to the relationship is to explore the behavior of the unstable circular photon orbit near the VdW-like phase transition point. The first work on this subject was recently completed in Ref. \cite{WeiLiuLiu} in the background of a $d$-dimensional charged AdS black hole. Two quantities, the radius and the minimum impact parameter of the unstable circular orbits of the null geodesics were numerically solved. And two main novel results were presented. One is that when the VdW-like phase transition occurs, both the radius and the minimum impact parameter demonstrate non-monotonous behaviors as the temperature increases, which is just like the behavior of the isothermal or isobaric lines of the VdW fluid indicating phase transition. While when the temperature and pressure are above the second-order critical point, the radius and the minimum impact parameter only show monotonous behaviors as a function of the temperature, and thus no phase transition is revealed. The other one is that the differences in the radius and the minimum impact parameter for the small and large black hole phases have a critical exponent of $\frac{1}{2}$. And their behaviors confirm that the differences can serve as the order parameters describing the black hole phase transition. Considering these two aspects, one can claim that there indeed exists a relationship between the gravity and the thermodynamic phase transition. The behavior of the unstable circular photon orbit carries the phase transition information.

Besides the charged AdS black hole, the rotating AdS black hole also displays the small-large black hole phase transition, so whether this relationship holds for the rotating AdS black hole is of great value. The aim of this paper is to examine the relationship between the unstable circular photon orbit and the phase transition for the rotating Kerr-AdS black hole. This has a great meaning on studying the black hole thermodynamics in the rotating spacetime.

This work is organized as follows. In Sec. \ref{bhg}, we give a brief review of the thermodynamics and phase transition for the rotating Kerr-AdS black hole. The coexistence curves and metastable curves are introduced. In Sec. \ref{geod}, the geodesics is studied in the equatorial plane. Employing the effective potential, the equations for the unstable circular photon orbits are obtained and the orbits are found to closely depend on the pressure of the black hole. The relationship between the unstable circular photon orbit and the thermodynamic phase transition is examined in Sec. \ref{relation} from three aspects: i) isobar and isotherm, ii) order parameter and critical exponent, and iii) metastable curves. Finally, the conclusions and discussions are presented in Sec. \ref{Conclusion}.

\section{Thermodynamics for Kerr-AdS black holes}
\label{bhg}

In this section, we would like to give a brief review of the thermodynamics and phase transition for the Kerr-AdS black holes.

The line element described the Kerr-AdS black hole is
\begin{eqnarray}
 ds^{2}=-\frac{\Delta}{\rho^{2}}\bigg(dt-\frac{a\sin^{2}\theta}{\Xi}d\varphi\bigg)^{2}
        +\frac{\rho^{2}}{\Delta}dr^{2}+\frac{\rho^{2}}{1-a^{2}/l^{2}\cos^{2}\theta}d\theta^{2}\nonumber\\
        +\frac{(1-a^{2}/l^{2}\cos^{2}\theta)\sin^{2}\theta}{\rho^{2}}\bigg(adt-\frac{r^{2}+a^{2}}{\Xi}d\varphi\bigg)^{2},
\end{eqnarray}
with the metric functions given by
\begin{eqnarray}
 \Delta&=&(r^{2}+a^{2})(1+r^{2}/l^{2})-2mr,\\
 \rho^{2}&=&r^{2}+a^{2}\cos^{2}\theta,\quad
 \Xi=1-\frac{a^{2}}{l^{2}},
\end{eqnarray}
where $m$ and $a$ are the black hole mass and spin parameters. The AdS radius $l$ related to the cosmological constant $\Lambda$ is interpreted as a pressure
\begin{equation}
 P=-\frac{\Lambda}{8\pi}=\frac{3}{8\pi l^{2}}.
\end{equation}
When taking $\Lambda$=0 or $l=\infty$, this black hole solution will reduce to the Kerr black hole, and only two black hole branches exist. One is stable and the other one is unstable. Thus, there is no the VdW-like phase transition. Taking $a$=0, it describes the Schwarzschild-AdS black hole, which has one unstable small branch and one stable large branch. Upon further considering of the thermal gas phase in the space, there will exist a Hawking-Page phase transition between the thermal gas and large black holes. The temperature, entropy, and angular velocity are
\begin{eqnarray}
 T&=&\frac{r_{\rm h}}{4\pi(r_{\rm h}^{2}+a^{2})}
   \left(1+\frac{a^{2}}{l^{2}}+3\frac{r_{\rm h}^{2}}{l^{2}}-\frac{a^{2}}{r_{\rm h}^{2}}\right),\\
 S&=&\frac{\pi (r_{\rm h}^{2}+a^{2})}{\Xi},\quad \Omega=\frac{a\Xi}{r_{\rm h}^{2}+a^{2}}+\frac{a}{l^{2}},
\end{eqnarray}
where $r_{\rm h}$ is the horizon radius of the black hole and can be obtained by solving $\Delta(r_{\rm h})=0$. The black hole mass $M$ and angular momentum $J$ are
\begin{eqnarray}
 M=\frac{m}{\Xi^{2}},\quad J=\frac{am}{\Xi^{2}}.
\end{eqnarray}
Considering that the cosmological constant is treated as the pressure for the black hole system, the black hole mass $M$ needs to be identified as the enthalpy $H$. Thus the Gibbs free energy $G=M-TS$ is given by
\begin{eqnarray}_{\rm h}
 G=\frac{a^4 \left(r_{\rm h}^2-l^{2}\right)+a^2\left(3l^{4}
   +2l^{2}r_{\rm h}^2+3r_{\rm h}^4\right)+l^{2} r_{\rm h}^2
   \left(l^{2}-r_{\rm h}^2\right)}{4 r_{\rm h}\left(a^2-l^{2}\right)^2}.\label{Gibbsfree}
\end{eqnarray}
After a simple calculation, the enthalpy, temperature, Gibbs free energy, and thermodynamic volume can be reexpressed in the following forms
\begin{eqnarray}
 H&=&\sqrt{\frac{(8PS+3)(12\pi^{2}J^{2}+S^{2}(8PS+3))}{36\pi S}},\\
 T&=&\frac{S^2 \left(64 P^2S^2+32PS+3\right)
     -12\pi^2 J^2}{4\sqrt{\pi} S^{3/2} \sqrt{8 P S+3} \sqrt{12 \pi^2 J^2+S^2(8 P S+3)}},\label{TT}\\
 G&=&\frac{12\pi^2 J^2(16PS+9)-64 P^2 S^4+9S^2}{12\sqrt{\pi}\sqrt{S}\sqrt{8PS+3}
    \sqrt{12\pi^2J^2+S^2 (8PS+3)}},\label{Gibbsfree}\\
 V&=&\frac{4\sqrt{S}\left(6\pi^2J^2+S^2(8PS+3)\right)}{3\sqrt{\pi}\sqrt{8PS+3}
   \sqrt{12\pi^2J^2+S^2(8PS+3)}}.\label{VV}
\end{eqnarray}
It was shown in Ref. \cite{Gunasekaran} that this Kerr-AdS black hole admits a small-large black hole phase transition. And similar to the VdW fluid, there also exists a critical point. However, it is very hard to obtain the critical point for this black hole by using the conditions $(\partial_{V}P)_{T,J}=(\partial_{V,V}P)_{T,J}=0$. In the small $J$ limit, the approximate critical point was obtained. On the other hand, we reexamined the thermodynamics for the Kerr-AdS black holes, and obtained the analytical critical point by employing the conditions $(\partial_{S}T)_{P,J}=(\partial_{S,S}T)_{P,J}=0$ \cite{Wei1}. Considering that the analytical critical point is in a complicated form, we just show their numerical result
\begin{eqnarray}
 P_{\rm c} \simeq\frac{0.002857}{J},\quad
 T_{\rm c} \simeq\frac{0.041749}{\sqrt{J}},\quad
 V_{\rm c} \simeq 115.796503 \sqrt[3]{J},\quad
 S_{\rm c} \simeq 28.718873J.
\end{eqnarray}
Moreover, the small-large black holes phase transition was also studied. The parametrized form of the coexistence curve in the reduced $\tilde{P}$-$\tilde{T}$ diagram is
\begin{eqnarray}
 \tilde{P}&=&0.718728 \tilde{T}^2 + 0.188963 \tilde{T}^3 + 0.0598479 \tilde{T}^4 + 0.0272726 \tilde{T}^5\nonumber\\
  &-&0.00597058 \tilde{T}^6 + 0.0203683 \tilde{T}^7 - 0.0159124 \tilde{T}^8 +
 0.00897128 \tilde{T}^9 - 0.00227244 \tilde{T}^{10},  \label{coex4}
\end{eqnarray}
with the reduced parameters being defined as
\begin{eqnarray}
 \tilde{P}=\frac{P}{P_{\rm c}},\quad \tilde{T}=\frac{T}{T_{\rm c}}.
\end{eqnarray}
Note that other reduced thermodynamic quantities are also defined like the temperature and pressure. Based on this parametrized form of the coexistence curve, we can easily obtain the changes of these physical quantities when the phase transition occurs.

\begin{figure}
\center{\subfigure[]{\label{HorizonRad}
\includegraphics[width=7cm]{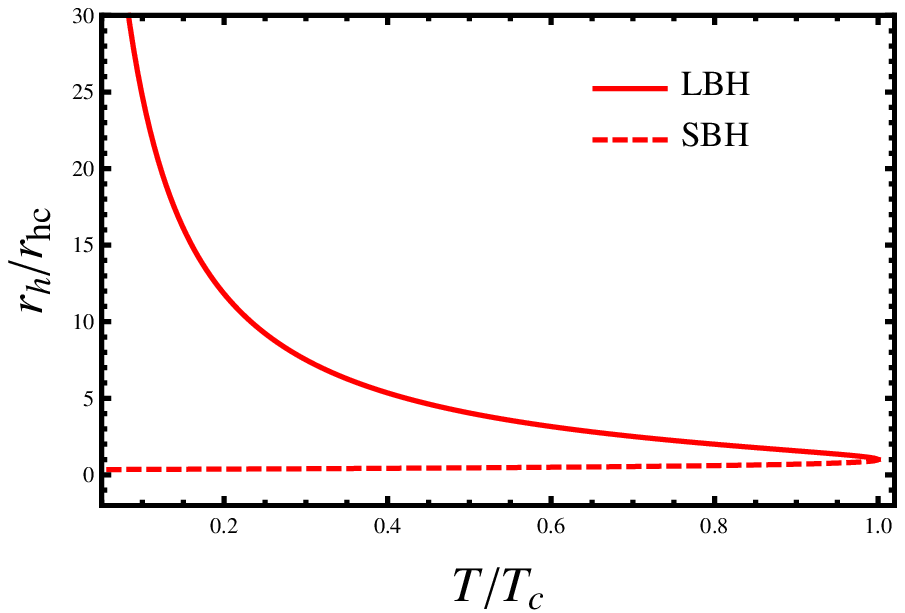}}
\subfigure[]{\label{HorizonChan}
\includegraphics[width=7cm]{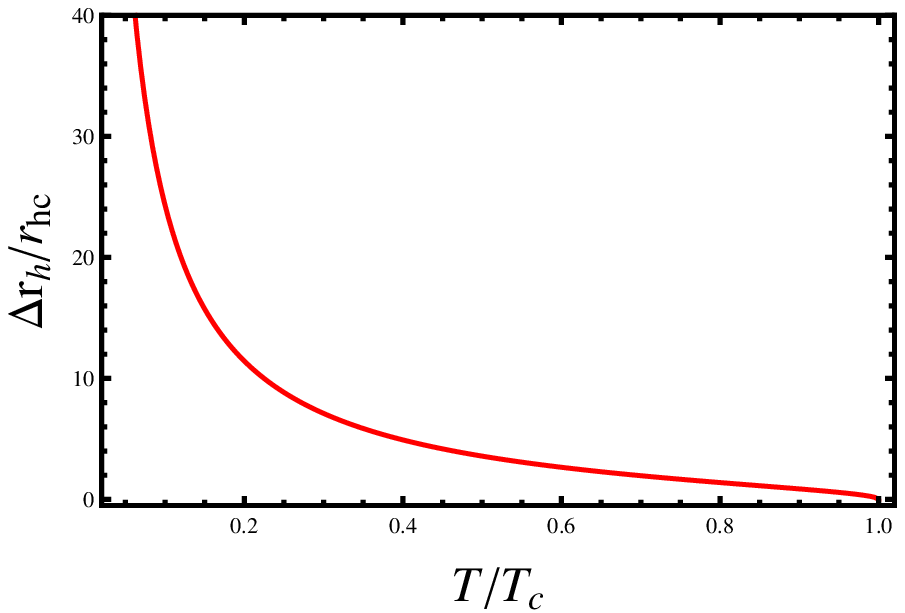}}}
\caption{(a) The horizon radii as a function of the phase transition temperature. The top and bottom curves are for the coexistence large and small black holes, respectively. (b) The difference of the horizon radii as a function of the phase transition temperature. }\label{pHorizonChan}
\end{figure}

On the other hand, the radius of the horizon denotes the size of the black hole. So it is interesting to examine the change of the radius among the phase transition. Here we express the radius $r_{\rm h}$ as
\begin{eqnarray}
 r_{\rm h}=\left(\frac{S^{3}(3+8PS)}{12\pi^{3}J^{2}+\pi S^{2}(3+8PS)}\right)^{\frac{1}{2}}.
\end{eqnarray}
At the critical point, we have the critical radius
\begin{eqnarray}
 r_{\rm hc}=2.965810\sqrt{J}.
\end{eqnarray}
In Fig. \ref{HorizonRad}, we show the horizon radius of the black hole as a function of the phase transition temperature. It is clearly that, for a given temperature, the two coexistent black holes have different horizon radii, which correspond to the small and large black holes. And the difference between them is also plotted in Fig. \ref{HorizonChan}. Moreover, with the increasing of the phase transition temperature, the difference of the horizon radii between these two coexistent black holes decreases, and at the critical point, the two radii coincide. From the other side, this phase transition characterized by the size change of the black holes is in fact a result of the change of the black hole microstructure \cite{Weisw}.

\begin{figure}
\center{\subfigure[]{\label{Phasert}
\includegraphics[width=7cm]{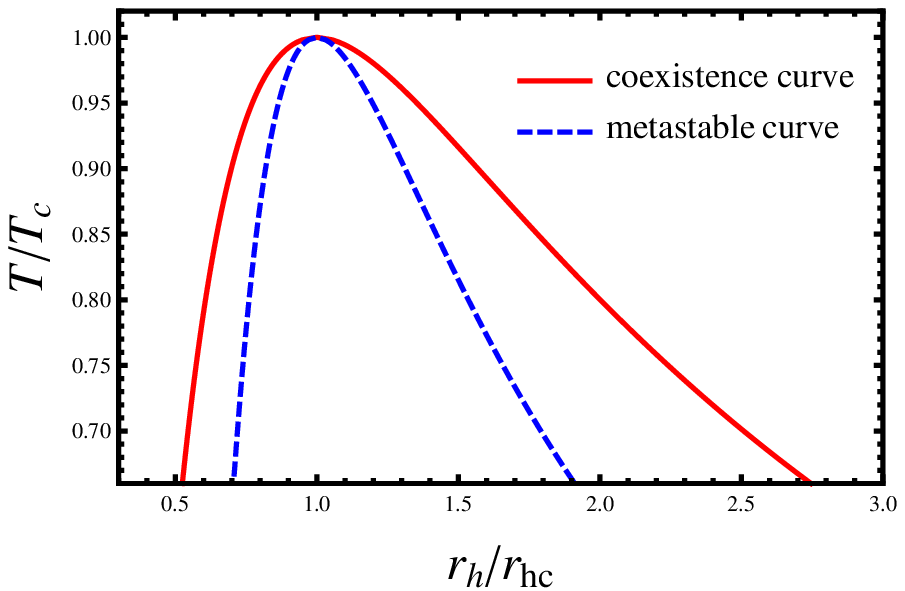}}
\subfigure[]{\label{Phasept}
\includegraphics[width=7cm]{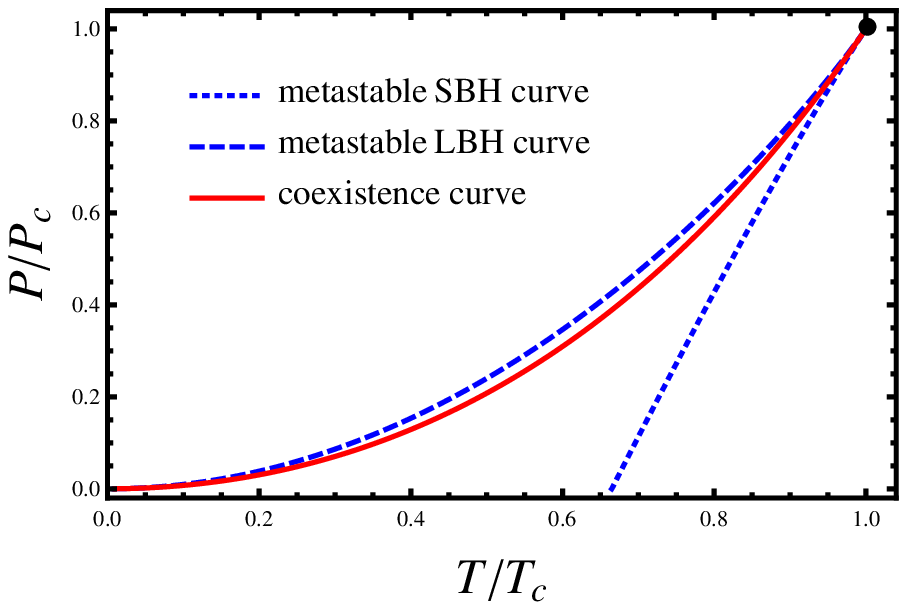}}}
\caption{(a) $\tilde{T}$-$\tilde{r}_{\rm h}$ phase diagram. (b) $\tilde{P}$-$\tilde{T}$ phase diagram.}\label{pPhasept}
\end{figure}

For the sake of clarity, we show the phase diagram in Fig. \ref{pPhasept}. In Fig. \ref{Phasert}, the phase structure is displayed in the reduced $\tilde{T}$-$\tilde{r}_{\rm h}$ diagram. The region below the red solid coexistence curve is a coexistence phase of small and large black holes. And the regions located between the coexistence curve and metastable curve (blue dashed line) are for the superheated small black hole phase and supercooled large black hole phase, respectively, which are two metastable phases of the black hole system. The $\tilde{P}$-$\tilde{T}$ phase diagram is described in Fig. \ref{Phasept}. The black dot denotes the critical point. In particular, the metastable small and large black hole curves are clearly shown. These curves are determined by different conditions. For example, for the coexistence curve, it satisfies
\begin{eqnarray}
 (\Delta G)_{P, T}=0.
\end{eqnarray}
While the metastable curve satisfies
\begin{eqnarray}
 (\partial_{S}T)_{P}=0,\quad \texttt{or}\quad
 (\partial_{V}P)_{T}=0,
\end{eqnarray}
which means that the metastable curve is the extremal point in the $T$-$S$ diagram or $P$-$V$ for a fixed pressure or temperature, respectively.

From above, we can find that the physical properties of the black holes have a close link with the size of its horizon, and thus, there may exist some interesting phenomena when the black hole system encounters a phase transition. In the following, we will discuss the behavior of the null geodesics near the thermodynamic phase transition.

It is worthwhile to point out that the coexistence small and large black hole curves are the same ones in the $\tilde{P}$-$\tilde{T}$ diagram, while the metastable small and large black hole curves are two different ones in that diagram. So in this $\tilde{P}$-$\tilde{T}$ diagram, the coexistence curves and metastable curves behave very differently.

\section{Geodesics for Kerr-AdS black holes}
\label{geod}

In this section, we would like to consider the geodesics of a test particle in the Kerr-AdS black hole background. For simplicity, we limit the orbits of the particle in the equatorial plane with $\theta=\frac{\pi}{2}$.

In the equatorial plane, the reduced metric can be expressed as
\begin{eqnarray}
 ds^{2}=-Adt^{2}+Bdr^{2}+Cd\phi^{2}-2 D dtd\varphi,
\end{eqnarray}
where these metric functions are given by
\begin{eqnarray}
 A(r)&=&\frac{\Delta-a^{2}}{r^{2}},\quad
 B(r)=\frac{r^{2}}{\Delta},\nonumber\\
 C(r)&=&\frac{a^{4}+r^{4}+a^{2}(2r^{2}-\Delta)}{r^{2}\Xi^{2}},\quad
 D(r)=\frac{a(a^{2}+r^{2}-\Delta)}{r^{2}\Xi^{2}}.
\end{eqnarray}
This Kerr-AdS black hole background admits two Killing fields $\partial_{t}$ and $\partial_{\varphi}$. So there are two constants $E$ and $L$, which are the conservation of energy and orbital angular momentum per unit mass of the motion. Combining with the momentum $p_{\mu}=g_{\mu\nu}\dot{x}^{\nu}$, we have
\begin{eqnarray}
 -E&=&p_{t}=-A\dot{t}-D\dot{\varphi},\\
 L&=&p_{\varphi}=-D\dot{t}+C\dot{\varphi}.
\end{eqnarray}
Solving them, we can obtain the $t$-motion and $\varphi$-motion
\begin{eqnarray}
 \dot{t}&=&\frac{EC-L D}{D^{2}+AC},\\
 \dot{\varphi}&=&\frac{ED+L A}{D^{2}+AC}.
\end{eqnarray}
Considering the normalizing condition $g_{\mu\nu}\dot{x}^{\mu}\dot{x}^{\nu}=-\delta^{2}$, one can get the $r$-motion,
\begin{eqnarray}
 \dot{r}^{2}=\frac{CE^{2}-L(2DE+AL)}{B(D^{2}+AC)}-\frac{\delta^{2}}{B},
\end{eqnarray}
where $\delta^{2}=0$ and $1$ for null geodesics and timelike geodesics, respectively. Moreover, we can rewrite the $r$-motion as
\begin{eqnarray}
 \dot{r}^{2}+V_{\rm eff}=0,
\end{eqnarray}
with the effective potential given by $V_{\rm eff}=\frac{L(2DE+AL)-CE^{2}}{B(D^{2}+AC)}+\frac{\delta^{2}}{B}$. For simplicity, we set $E$=1, which is equivalent to set $L\rightarrow LE$, $\delta\rightarrow \delta E$, and $V_{\rm eff}\rightarrow V_{\rm eff}/E^{2}$.

\begin{figure}
\center{\subfigure[]{\label{EffectPL2}
\includegraphics[width=7cm]{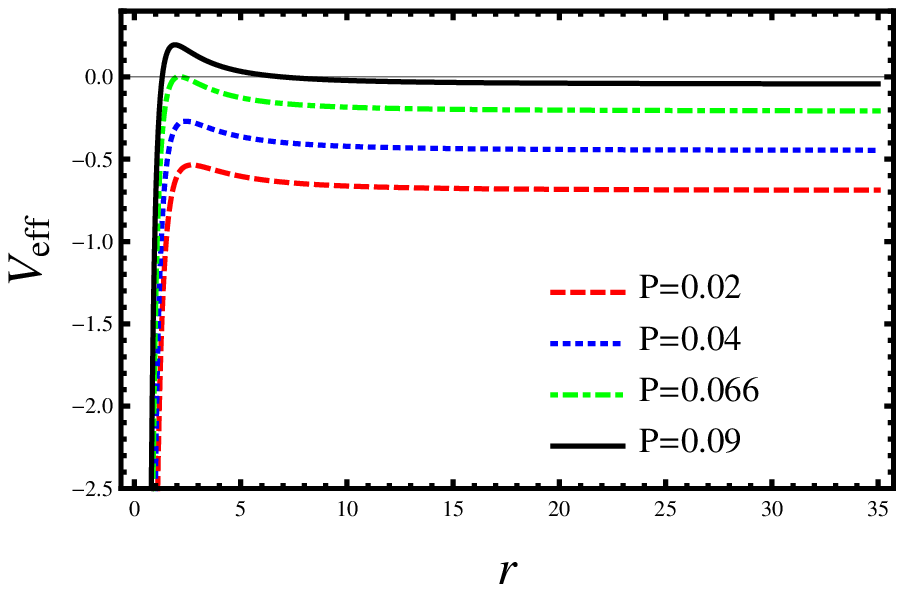}}
\subfigure[]{\label{EffectPL5}
\includegraphics[width=7cm]{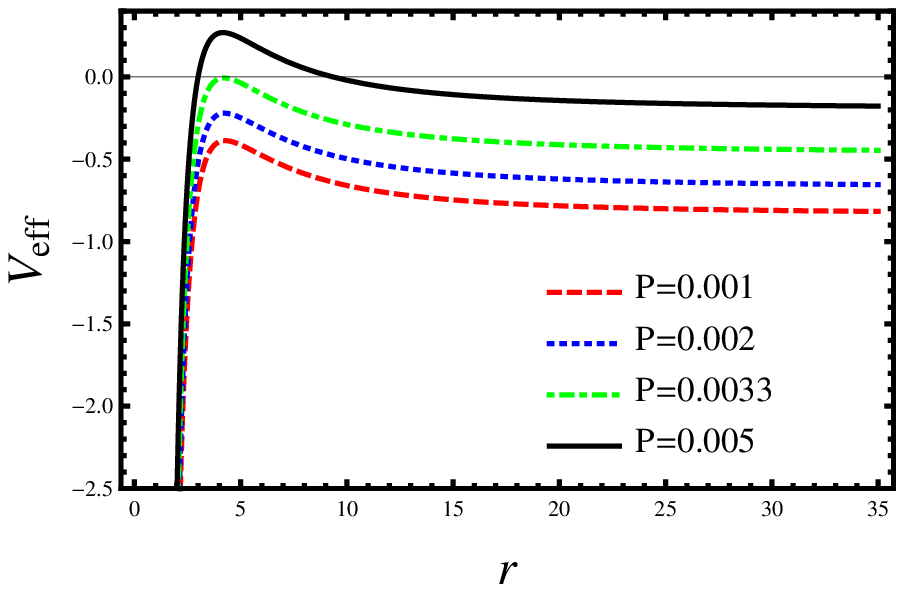}}}
\caption{Behavior of the effective potential with fixed $S$=1 and $J$=1. (a) $L$=2 and $P$=0.02, 0.04, 0.066, 0.09 from bottom to top. (b) $L$=5 and $P$=0.001, 0.002, 0.0033, 0.005 from bottom to top.}\label{pEffectPL5}
\end{figure}

Here, we show the behavior of the effective potential for the photon in Fig. \ref{pEffectPL5} for different values of the parameters. Since $\dot{r}^{2}>0$, these ranges of the negative effective potential are allowed for the photon to travel. For fixed $S$, $J$, and $L$, we can find that there exists a peak. With the increase of the pressure $P$, the peak increases and approaches to zero at some certain angular momentum $L$. Then further increasing $P$, the peak will be positive, and thus it will prevent the photon from outside falling into the black hole.

The interesting phenomena occur at the point where the peak just achieves zero. Under this situation, the photon has a vanished radial velocity while with a nonvanished transverse velocity. Thus, the photon will circle the black hole one loop by one loop. Such an unstable circular photon orbit is determined by
\begin{eqnarray}
 V_{\rm eff}(r_{\rm co}, L_{\rm co})=0, \quad
 \partial_{r}V_{\rm eff}(r_{\rm co}, L_{\rm co})=0.
\end{eqnarray}
Solving these conditions, we can obtain the radius $r_{\rm co}$ and the angular momentum $L_{\rm co}$ for the circular orbit. Plugging the effective potential into the above two conditions, we have
\begin{eqnarray}
 &&L_{\rm co}=\frac{-D(r_{\rm co})+\sqrt{A(r_{\rm co})C(r_{\rm co})+D(r_{\rm co})^{2}}}{A(r_{\rm co})},\label{c1}\\
 &&2L_{\rm co}\big(A(r_{\rm co})D'(r_{\rm co})-A'(r_{\rm co})D(r_{\rm co})\big)-\big(A(r_{\rm co})C'(r_{\rm co})-A'(r_{\rm co})C(r_{\rm co})\big)=0.\label{c2}
\end{eqnarray}
The prime indicates the derivative with respect to $r$.

Next, we would like to study the behaviors of the radius $r_{\rm co}$ and the angular momentum $L_{\rm co}$ for the circular orbit as functions of the temperature and pressure. Because in the reduced parameter space, these thermodynamic quantities are angular momentum $J$ independent, we also expect to explore $r_{\rm co}$ and $L_{\rm co}$ in the reduced parameter space. At the critical point of the phase transition, we find the critical values
\begin{eqnarray}
 r_{\rm coc}\approx4.518692\sqrt{J},\quad
 L_{\rm coc}\approx5.335057\sqrt{J}.
\end{eqnarray}

\section{Unstable circular photon orbit and thermodynamic phase transition}
\label{relation}

In this section, we would like to explore the quantities, the radius and the angular momentum, of the unstable circular photon orbit as the function of the thermodynamic parameter. We expect there exist some novel phenomena of the circular orbit when the thermodynamic phase transition takes place. These results will provide us insight on studying the relationship between the gravity and thermodynamics of the black hole system. Before carrying out the calculation and analysis, we note here that i) a thermodynamic first-order phase transition takes place for $\tilde{T}<1$ and $\tilde{P}<1$; ii) a second-order phase transition occurs at $\tilde{T}=1$ and $\tilde{P}=1$; iii) no phase transition exists when $\tilde{T}$ and $\tilde{P}$ are beyond 1.

\subsection{Isobar and isotherm}

Given black hole parameters, we can solve the radius $r_{\rm co}$ and the angular momentum $L_{\rm co}$ for the circular orbit by employing Eqs. (\ref{c1}) and (\ref{c2}). First, we aim to study the behaviors of $r_{\rm co}$ and $L_{\rm co}$ along the isobar and isotherm.

\begin{figure}
\center{\subfigure[]{\label{TRc}
\includegraphics[width=7cm]{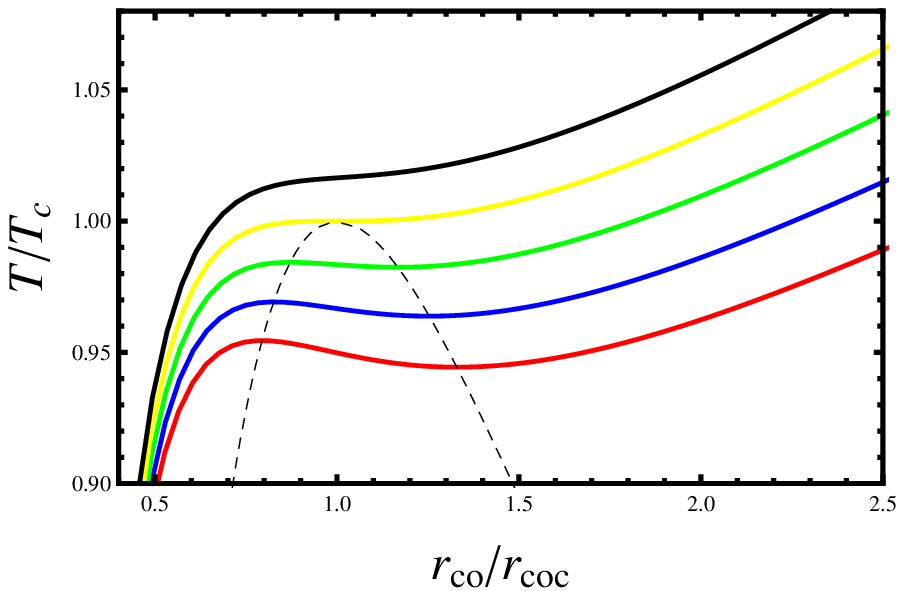}}
\subfigure[]{\label{TLc}
\includegraphics[width=7cm]{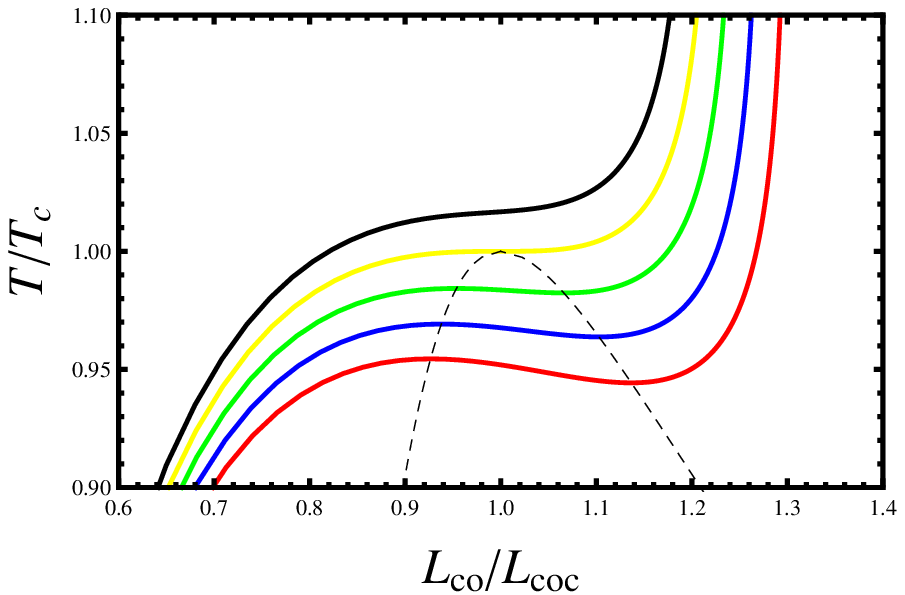}}}
\caption{Phase transition temperature as a function of the radius and the angular momentum of the circular orbit. (a) $\tilde{T}$-$\tilde{r}_{\rm co}$. (b) $\tilde{T}$-$\tilde{L}_{\rm co}$. The pressure $\tilde{P}$=0.88, 0.92, 0.96, 1, and 1.04 from bottom to top. And the black dashed curves correspond to the extremal points.}\label{pTLc}
\end{figure}

\begin{figure}
\center{\subfigure[]{\label{PRc}
\includegraphics[width=7cm]{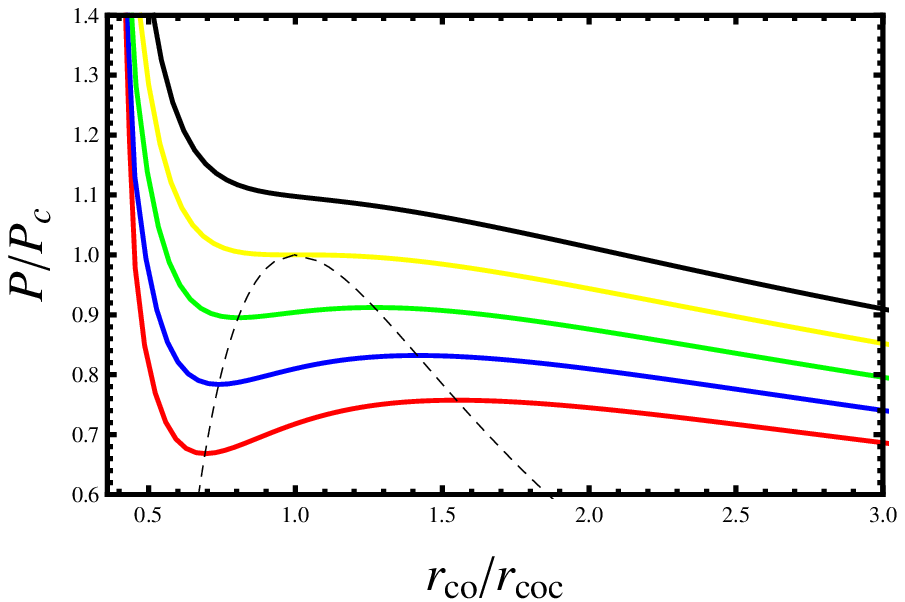}}
\subfigure[]{\label{PLc}
\includegraphics[width=7cm]{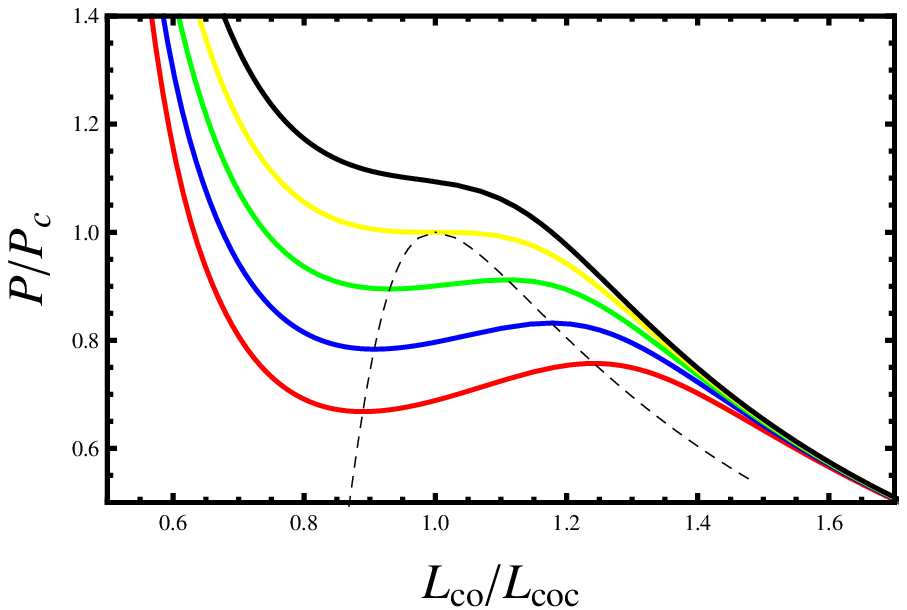}}}
\caption{Phase transition pressure as a function of the radius and the angular momentum of the circular orbit. (a) $\tilde{P}$-$\tilde{r}_{\rm co}$. (b) $\tilde{P}$-$\tilde{L}_{\rm co}$. The temperature $\tilde{T}$=0.88, 0.92, 0.96, 1, and 1.04 from bottom to top. And the black dashed curves correspond to the extremal points.}\label{pPLc}
\end{figure}

After a simple calculation, we plot in Fig.~\ref{pTLc} the reduced temperature $\tilde{T}$ as a function of the reduced radius $\tilde{r}_{\rm co}$ and angular momentum $\tilde{L}_{\rm co}$ of the circular orbit for a fixed pressure $\tilde{P}$=0.88, 0.92, 0.96, 1, and 1.04 from the bottom to top. From the figures, we can find that a low temperature is related to a small orbit radius and angular momentum. While for a high temperature, both the radius and angular momentum take large values. Among these two cases, it has an interesting behavior at the intermediate temperature for different pressures. We summarize as follows. For a low pressure $\tilde{P}<1$, there are two extremal points. This case is quite similar to the isobaric lines in the $\tilde{T}$-$\tilde{S}$ diagram. With the increase of $\tilde{P}$, these two extremal points get close and coincide at $\tilde{P}=1$. Further increasing the pressure such that $\tilde{P}>1$, no extremal point exists and the temperature is a monotonically increasing function of $\tilde{r}_{\rm co}$ or $\tilde{L}_{\rm co}$. The similar conclusion can also be found in the $\tilde{P}$-$\tilde{r}_{\rm co}$ diagram and the $\tilde{P}$-$\tilde{L}_{\rm co}$ diagram; see Fig.~\ref{pPLc}. According to this feature of the radius and angular momentum of the circular orbit, one can directly read out whether there exists a thermodynamic phase transition by counting the number of the extremal points. For example, the existence of two extremal points implies a first-order phase transition, and one extremal point corresponds to a second-order phase transition. While no extreme point means no phase transition occurs. It is worthwhile to note that the thermodynamic phase transition point can be determined by constructing the equal area law. However, it is not sure that this equal area law is held in the cases of Figs.~\ref{pTLc} and \ref{pPLc}.

In summary, the behaviors of $r_{\rm co}$ and $L_{\rm co}$ confirm the same phase transition information as that from the thermodynamic side. So there indeed exists a relationship between the circular orbit and the thermodynamic phase transition.

\subsection{Order parameter and critical exponent}

From the Gibbs free energy (\ref{Gibbsfree}), we can determine the thermodynamic phase transition point. And the result has been given in Sec. \ref{bhg}. Here, we would like to study the differences of the radius $r_{\rm co}$ and the angular momentum $L_{\rm co}$ along the coexistence curve.

\begin{figure}
\center{\subfigure[]{\label{RcT}
\includegraphics[width=7cm]{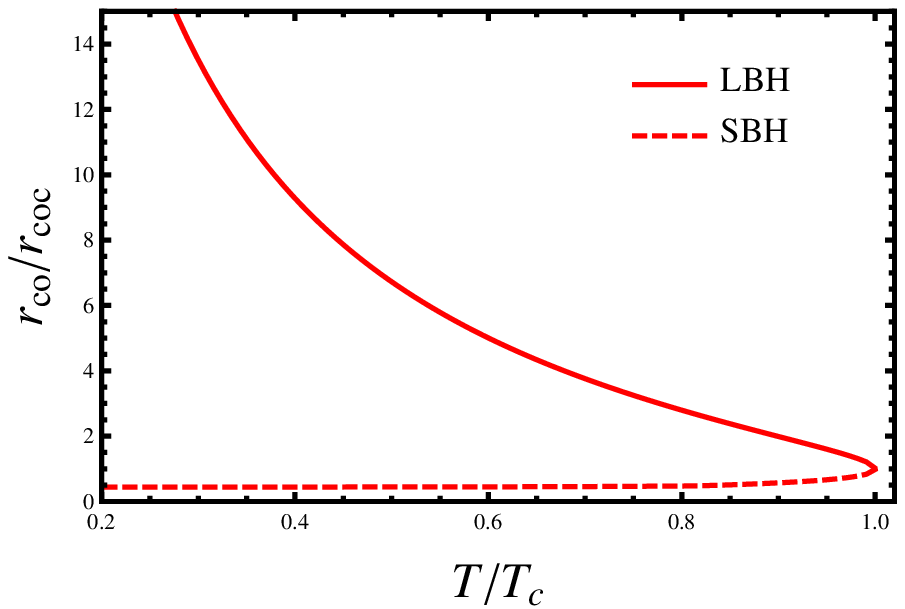}}
\subfigure[]{\label{DeltaRcT}
\includegraphics[width=7cm]{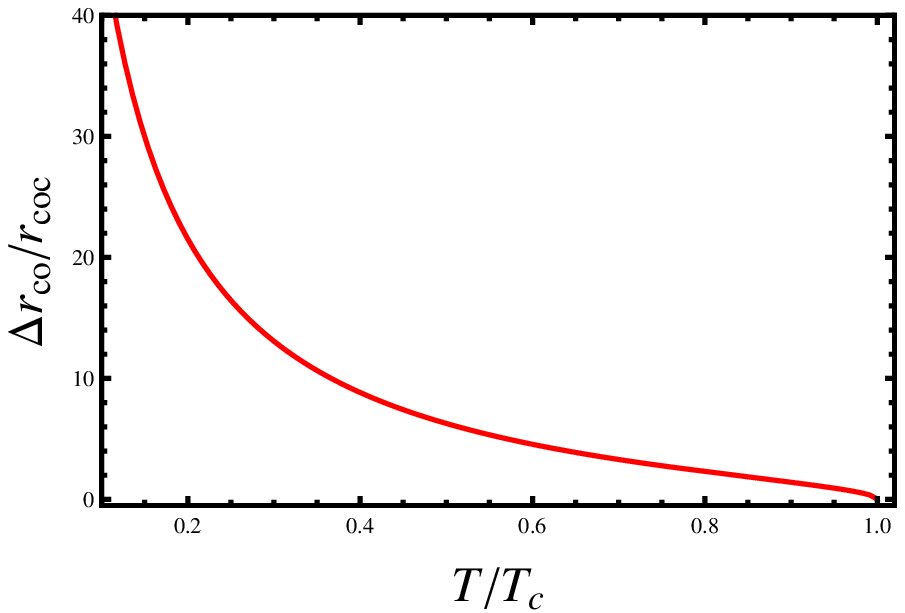}}}
\caption{(a) The radius of the circular orbit for the coexistence small and large black holes. (b) The difference of the radii of the circular orbit.}\label{pDeltaRcT}
\end{figure}

\begin{figure}
\center{\subfigure[]{\label{LcT}
\includegraphics[width=7cm]{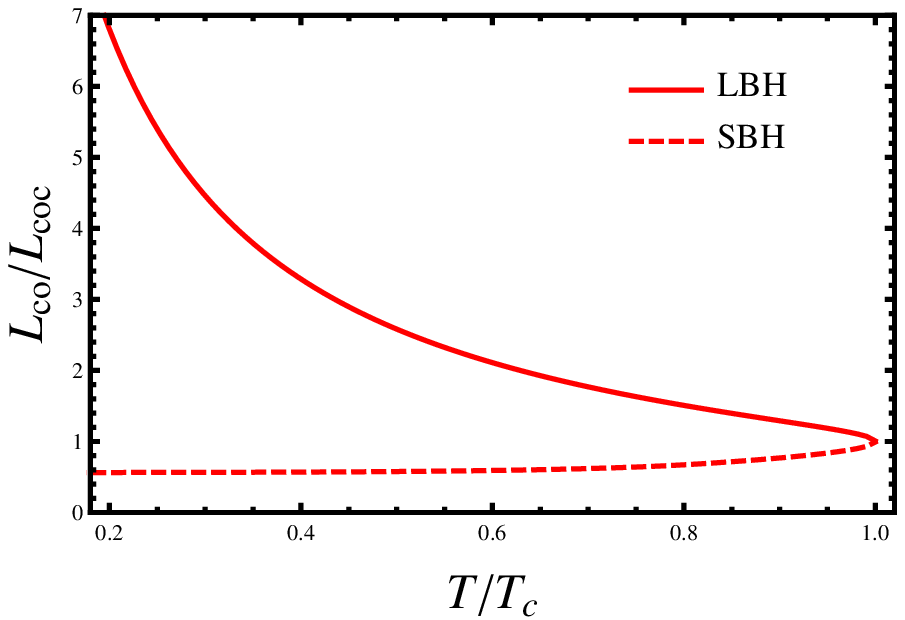}}
\subfigure[]{\label{DeltaLcT}
\includegraphics[width=7cm]{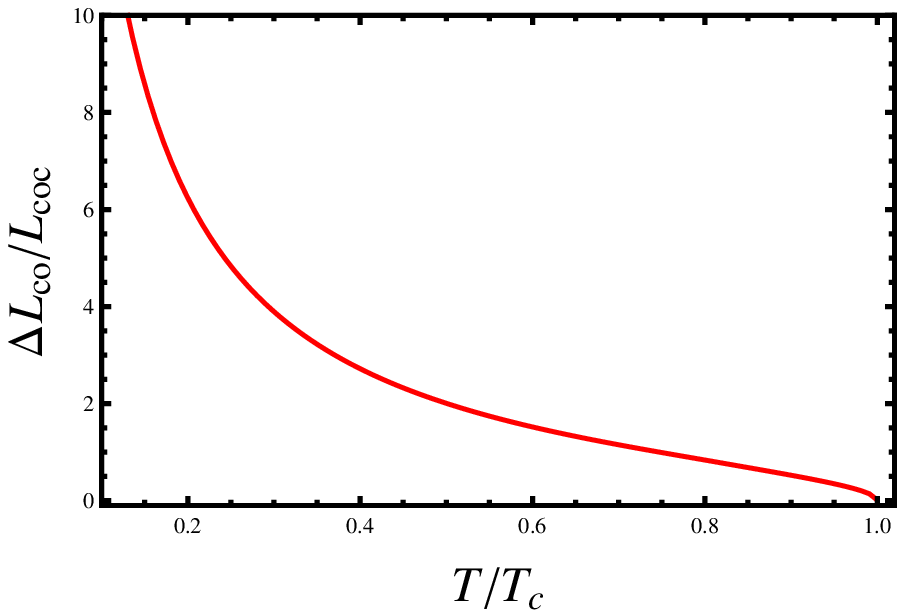}}}
\caption{(a) The angular momentum of the circular orbit for coexistence small and large black holes. (b) The difference of the angular momenta of the circular orbit.}\label{pDeltaLcT}
\end{figure}

The shape of the $\tilde{r}_{\rm co}$ is plotted as a function of the phase transition temperature $\tilde{T}$ in Fig. \ref{RcT}. From it, it is easy to find that with the increase of temperature, the radius $\tilde{r}_{\rm co}$ decreases for the coexistence large black hole, while increases for the coexistence small black hole. At the critical point $\tilde{T}=1$, $\tilde{r}_{\rm co}$ approaches to the same value for the coexistence small and large black holes. We also present the difference $\Delta r_{\rm co}/r_{\rm coc}$ between them in Fig. \ref{DeltaRcT}. Quite interestingly, this difference decreases with the phase transition temperature, and tends to zero at the critical point. The angular momentum $\tilde{L}_{\rm co}$ for coexistence small and large black holes is also given in Fig. \ref{LcT}, and the difference $\Delta L_{\rm co}/L_{\rm coc}$ is displayed in Fig. \ref{DeltaLcT}. Its behavior is similar to that of the radius of the circular orbit.

\begin{figure}
\center{\subfigure[]{\label{DDrt}
\includegraphics[width=7cm]{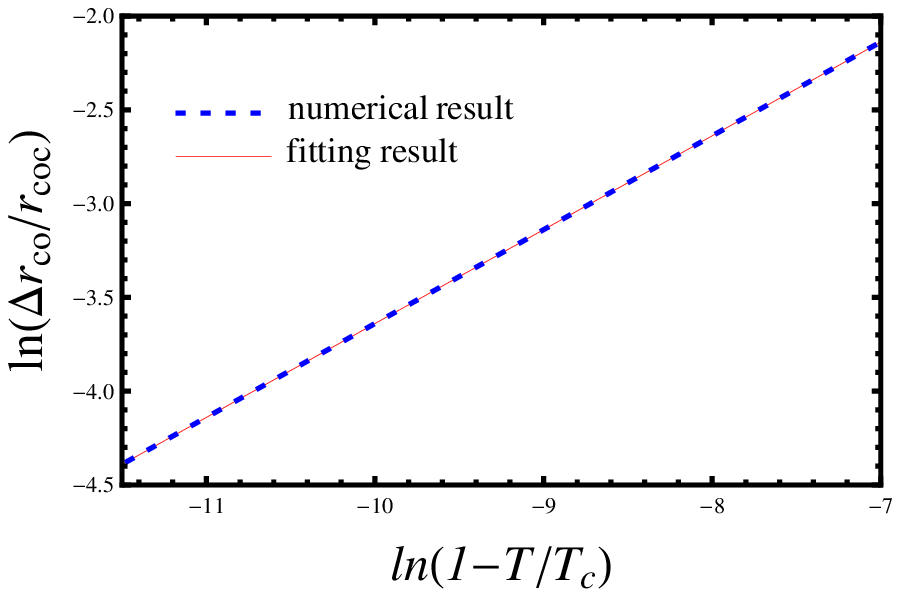}}
\subfigure[]{\label{DDLt}
\includegraphics[width=7cm]{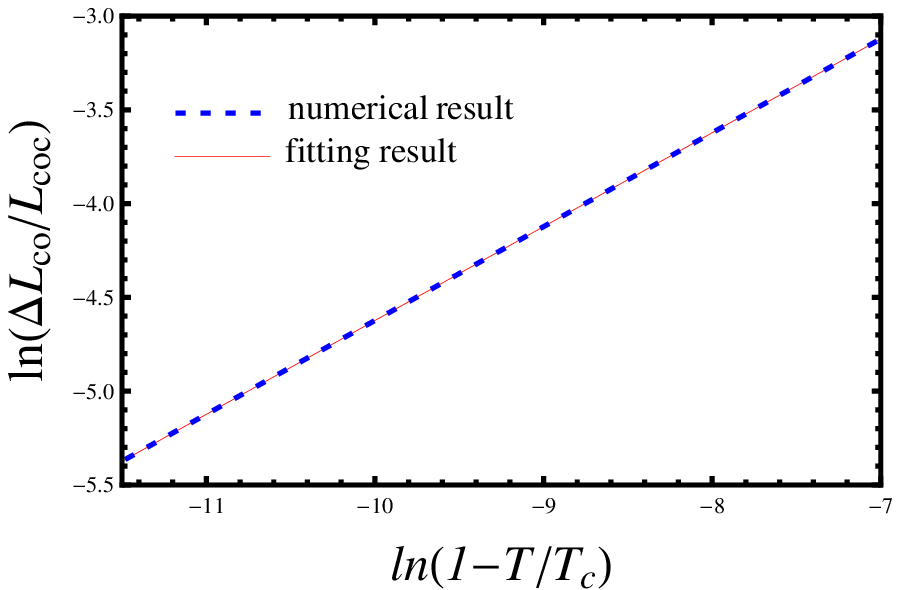}}}
\caption{Behaviors of the difference of the radius and angular momentum of the circular orbit near the critical point of the small-large black hole phase transition. Blue thick dashed lines are for the numerical results and red thin lines are our fitting results. (a) $\ln(\Delta r_{\rm co}/r_{\rm coc})$ vs $\ln(1-T/T_{\rm c})$. (b) $\ln(\Delta L_{\rm co}/L_{\rm coc})$ vs. $\ln(1-T/T_{\rm c})$.}\label{pDDLt}
\end{figure}

Remarkably, the behaviors of $\Delta r_{\rm co}/r_{\rm coc}$ and $\Delta L_{\rm co}/L_{\rm coc}$ are extremely like the order parameter. For example, they have a nonzero value at the first-order phase transition and vanish at the second-order phase transition. However, we also need to check their critical exponent $\delta$ at the critical point,
\begin{eqnarray}
 \Delta r_{\rm co}/r_{\rm coc},\;\Delta L_{\rm co}/L_{\rm coc}
  \sim a\times(1-\tilde{T})^{\delta}.
\end{eqnarray}
Taking the logarithm of them, we have
\begin{eqnarray}
 \ln(\Delta r_{\rm co}/r_{\rm coc}),\ln(\Delta L_{\rm co}/L_{\rm coc})\sim\delta\ln(1-\tilde{T})+\ln a,
\end{eqnarray}
which means that $\ln(\Delta r_{\rm co}/r_{\rm coc})$ and $\ln(\Delta L_{\rm co}/L_{\rm coc})$ are linear functions of $\ln(1-\tilde{T})$. By varying $\tilde{T}$ from 0.999 to 0.99999, we numerically obtain the result, which is described by the blue thick dashed lines in Fig. \ref{pDDLt}. By using the data, we find the following fitting results
\begin{eqnarray}
 \Delta r_{\rm co}/r_{\rm coc}&=&3.92146\times(1-\tilde{T})^{0.500638},\\
 \Delta L_{\rm co}/L_{\rm coc}&=&1.46624\times(1-\tilde{T})^{0.500626}.
\end{eqnarray}
Such fitting results confirm that the critical exponents for $\Delta r_{\rm co}/r_{\rm coc}$ and $\Delta L_{\rm co}/L_{\rm coc}$ equal to ${1}/{2}$, which are the same as that of the $d$-dimensional charged AdS black holes \cite{WeiLiuLiu}. Thus, $\Delta r_{\rm co}/r_{\rm coc}$ and $\Delta L_{\rm co}/L_{\rm coc}$ can serve as order parameters to describe the thermodynamic phase transition. Moreover, we also show the fitting result with the red thin solid lines in Fig. \ref{pDDLt}. Clearly, the numerical results and the fitting results highly consistent with each other, which, to some extent, confirms the critical behavior.

\subsection{Metastable curves}

As shown above, the metastable curves are also an important phenomenon in the black hole thermodynamics. Our result also shows that the radius and the angular momentum of the unstable circular photon orbit have the extremal points; see Figs.~\ref{pTLc} and \ref{pPLc}. So a natural conjecture is whether the extremal points of $\tilde{r}_{\rm co}$ and $\tilde{L}_{\rm co}$ coincide with the metastable curves in the $\tilde{P}$-$\tilde{T}$ diagram.

\begin{figure}
\center{\includegraphics[width=7cm]{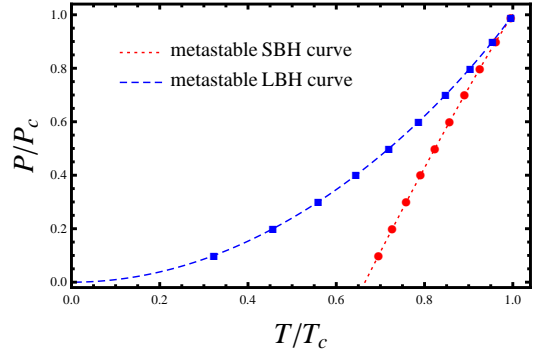}}
\caption{Extremal points in the $\tilde{P}$-$\tilde{T}$ diagram. Blue dashed curve and red dotted curve are the metastable large and small black hole curves from the thermodynamics, respectively. The blue square markers and red circle markers are the extremal points for the large and small black holes calculated from the circular orbit.}\label{pHorizon}
\end{figure}

When the reduced pressure $\tilde{P}$ takes values from 0.1 to 1, we first obtain the values of $\tilde{r}_{\rm co}$. Then combining the pressure and radius, we can derive the value for the temperature. Here we list the results in Fig.~\ref{pHorizon}. The blue square markers and the red circle markers are the extremal points for the large and small black holes, respectively. While the blue dashed curve and red dotted curve are the metastable large and small black hole curves shown in Fig.~\ref{Phasept}. From Fig.~\ref{pHorizon}, it is obvious to find that the extremal points of $\tilde{r}_{\rm co}$ well coincide with the metastable curves both for the small and large black holes. On the other hand, one can carry out the calculation by using $\tilde{l}_{\rm co}$, and we believe that the pressure and temperature obtained from the extremal points of $\tilde{l}_{\rm co}$ also agree with the metastable curves. So we will not carry out the corresponding calculation.

\begin{table}[h]
\begin{center}
\begin{tabular}{ccccccccc}
  \hline\hline
  $\tilde{P}$& $\tilde{T}^{\rm SBH}$& $\tilde{T}^{\rm SBH}_{\rm co}$& $\tilde{T}^{\rm LBH}$& $\tilde{T}^{\rm LBH}_{\rm co}$&$\tilde{r}_{\rm co}^{\rm SBH}$&$\tilde{r}_{\rm h}^{\rm SBH}$&$\tilde{r}_{\rm co}^{\rm LBH}$&$\tilde{r}_{\rm h}^{\rm LBH}$\\\hline
0.1&0.32297180294&0.32297180289&0.6958506&0.6958516&0.829541&0.718227&3.43559&3.97583\\
0.2&0.45653718258&0.45653718255&0.7271420&0.7271451&0.838743&0.729782&2.43409&2.80403\\
0.3&0.55869820508&0.55869820496&0.7588437&0.7588446&0.848247&0.742426&1.99005&2.27928\\
0.4&0.64439512755&0.64439512739&0.7910013&0.7910028&0.858463&0.756443&1.72439&1.96100\\
0.5&0.71936444356&0.71936444232&0.8236740&0.8236742&0.869030&0.772249&1.54171&1.73818\\
0.6&0.78649484399&0.78649484027&0.8569402&0.8569409&0.880714&0.790501&1.40496&1.56752\\
0.7&0.84743484306&0.84743483511&0.8909116&0.8909120&0.893885&0.812334&1.29601&1.42754\\
0.8&0.90315665001&0.90315664410&0.9257610&0.9257612&0.909776&0.840028&1.20404&1.30482\\
0.9&0.95413885419&0.95413882629&0.9618041&0.9618040&0.931817&0.879689&1.11994&1.18662\\
0.99&0.99574023027&0.99574014916&0.9959743&0.9959743&0.975095&0.957165&1.02973&1.04901
\\\hline\hline
\end{tabular}
\caption{Numerical values for the temperature $\tilde{T}$ and $\tilde{T}_{\rm co}$ corresponding to the metastable curves and the extremal point of $\tilde{r}_{\rm co}$, respectively, and the radii for the horizon and circular orbit of the small and large black holes.}\label{tab1}
\end{center}
\end{table}

For a detailed study, we list the numerical results in Table~\ref{tab1}. From the numerical data, it is clear that, for the small black hole, $\tilde{T}$ and $\tilde{T}_{\rm co}$ have a deviation of about $10^{-7}\sim 10^{-9}$. And for the large black hole, the deviation is about $10^{-7}$. Therefore, we can conclude that the temperature calculated from the extremal points of $\tilde{r}_{\rm co}$ highly agrees with that of the metastable curves. So our conjecture is appropriate. Note the circular orbits should be out of the black hole horizon if one has $r_{\rm co}>r_{\rm h}$. However, from the table, one may argue that the horizon of the large black hole has a larger radius than the circular orbit; for example $\tilde{r}_{\rm h}^{LBH}=3.9758> \tilde{r}_{\rm co}^{LBH}=3.43559$ for $\tilde{P}=0.1$. In fact, the radius has been reduced by its critical value. Considering this case, we have $r_{\rm co}=15.5244\sqrt{J}>r_{\rm h}=11.7915\sqrt{J}$ for $\tilde{P}=0.1$, which means the circular orbit locates outside of the horizon.

\section{Conclusions and discussions}
\label{Conclusion}

In this paper, we examined the relationship between the unstable circular photon orbit and the thermodynamic phase transition for the rotating Kerr-AdS black hole. This provides a good way to realize the link between the gravity and thermodynamics of the strong gravity.

First, we presented several thermodynamic features of the black hole phase transition, such as the coexistence curves and metastable curves. Since these curves denote the boundary of different stable or metastable phases, they are keys to study the black hole phase structure. Also, with the help of the parametrized form of the coexistence curve, we calculated the horizon radius for the coexistence small and large black holes, which can be used to probe the thermodynamic property for different black hole phases.

Second, we investigated the null geodesics of a test particle in the equatorial plane of the black hole background. By the effective potential, we obtained the radius and the angular momentum of the circular orbits. It was found that these two quantities have a close relationship with the black hole pressure.

Then we explored the relationship between the unstable circular photon orbit and the thermodynamic phase transition. It mainly contains three aspects:

i) Isobar and isotherm for the radius and the angular momentum of the circular orbit. We found that when the reduced pressure $\tilde{P}$ or temperature $\tilde{T}$ is less than 1, both the radius $\tilde{r}_{\rm co}$ and the angular momentum $\tilde{L}_{\rm co}$ of the circular orbit demonstrate that there are two extremal points. When $\tilde{P}=\tilde{T}=1$, the two extremal points coincide with each other. When the pressure and temperature are greater than 1, no extremal point exists. Thus, $\tilde{r}_{\rm co}$ and $\tilde{L}_{\rm co}$ will be two monotonically increasing functions of the pressure or temperature. The behaviors of the isobar and isotherm of $\tilde{r}_{\rm co}$ and $\tilde{L}_{\rm co}$ are consistent with that of the black hole thermodynamics. Thus, the thermodynamic phase transition information can be revealed from the behavior of the circular orbit.

ii) Order parameter and critical exponent. At the first-order thermodynamic phase transition, there are two coexistence small and large black hole branches. And each branch has a different $\tilde{r}_{\rm co}$ and $\tilde{L}_{\rm co}$ for the circular orbit. Our result shows that the differences $\Delta \tilde{r}_{\rm co}$ and $\Delta \tilde{L}_{\rm co}$ for these two branches can act as order parameters to describe the black hole phase transition. More importantly, at the critical point, the fitting result displays that both the differences $\Delta \tilde{r}_{\rm co}$ and $\Delta \tilde{L}_{\rm co}$ have a critical exponent of $\frac{1}{2}$.

iii) Metastable curves. Considering the calculation error, the temperature and pressure corresponding to the extremal points of $\tilde{r}_{\rm co}$ and $\tilde{L}_{\rm co}$ for the circular orbit completely agree with that of the metastable curves.

These results firmly enhance our conjecture that there exists the relationship between the geodesics and thermodynamic phase transition for the Kerr-AdS black holes. It further supports the links between the gravity and thermodynamics for the black holes. This work is worth generalizing to other black hole backgrounds. It also provides a possible way to investigate the strong gravitational effect from the thermodynamic side.

\section*{Acknowledgements}
This work was supported by the National Natural Science Foundation of China (Grants No. 11675064, No. 11875151, and  No. 11522541), and the Fundamental Research Funds for the Central Universities (Grants No. lzujbky-2018-k11, No. lzujbky-2017-it69, and No. lzujbky-2017-182).

\end{document}